\documentstyle[12pt]{article}
\begin{document}
\newcommand{\ket}[1] {\mbox{$ \vert #1 \rangle $}}
\newcommand{\bra}[1] {\mbox{$ \langle #1 \vert $}}
\newcommand{\bkn}[1] {\mbox{$ < #1 > $}}
\newcommand{\bk}[1] {\mbox{$ \langle #1 \rangle $}}
\newcommand{\scal}[2]{\mbox{$ \langle #1 \vert #2  \rangle $}}
\newcommand{\expect}[3] {\mbox{$ \bra{#1} #2 \ket{#3} $}}
\newcommand{\ki}{\mbox{$ \ket{\psi_i} $}}
\newcommand{\bi}{\mbox{$ \bra{\psi_i} $}}
\newcommand{\p} \prime
\newcommand{\e} \epsilon
\newcommand{\la} \lambda
\newcommand{\om} \omega   \newcommand{\Om} \Omega
\newcommand{\cc}{\mbox{$\cal C $}}
\newcommand{\w} {\hbox{ weak }}
\newcommand{\al} \alpha
\newcommand{\bt} \beta
\newcommand{\be} {\begin{equation}}
\newcommand{\ee} {\end{equation}}
\newcommand{\ba} {\begin{eqnarray}}
\newcommand{\ea} {\end{eqnarray}}

\def\lrD{\mathrel{{\cal D}\kern-1.em\raise1.75ex\hbox{$\leftrightarrow$}}}

\def\lr #1{\mathrel{#1\kern-1.25em\raise1.75ex\hbox{$\leftrightarrow$}}}

\overfullrule=0pt \def\sqr#1#2{{\vcenter{\vbox{\hrule height.#2pt
          \hbox{\vrule width.#2pt height#1pt \kern#1pt
           \vrule width.#2pt}
           \hrule height.#2pt}}}}
\def\square{\mathchoice\sqr68\sqr68\sqr{4.2}6\sqr{3}6} 
\def\lrpartial{\mathrel
{\partial\kern-.75em\raise1.75ex\hbox{$\leftrightarrow$}}}

\begin{flushright}
March 12th, 1998
\\\end{flushright}
\vskip 1. truecm
\vskip 1. truecm
\centerline{\LARGE\bf{The Background Field Approximation}}
\vskip 2 truemm
\centerline{\LARGE\bf{in (quantum) cosmology}}
\vskip 1. truecm
\vskip 1. truecm

\centerline{{\bf R. Parentani}}
\vskip 5 truemm

\centerline{
Laboratoire de Math\'ematiques et Physique Th\'eorique,
CNRS UPRES A 6083}
\centerline{Facult\' e des Sciences, Universit\'e de Tours, 37200 Tours, France}
\centerline{
e-mail: parenta@celfi.phys.univ-tours.fr}
\vskip 5 truemm
\vskip 1.5 truecm
\vskip 1.5 truecm

{\bf Abstract }\\
We analyze the Hamilton-Jacobi
action of gravity and matter in the limit 
where gravity is treated at the
 background field approximation.
The motivation 
is to clarify 
when and how the solutions of the Wheeler-DeWitt equation
lead to the Schr\"odinger equation 
in a given background.
To this end, we determine when and how the total action, solution 
of the constraint equations of General Relativity,
leads to the HJ action for matter in a given background.
This is achieved by comparing two neighboring solutions
differing slightly in their matter energy content.
To first order in the 
change 
of the 3-geometries, 
the change of the gravitational action 
equals the integral of the 
matter energy 
evaluated in the background geometry.
Higher order terms are governed by the ``susceptibility'' of the
geometry.
These classical properties 
also
apply to quantum cosmology
since the conditions which legitimize the use of WKB
gravitational waves are concomitant with those
governing the validity of the 
background field approximation.

\vfill \newpage

\section{Introduction}

There is a question that haunts all works
in which 
both quantum matter effects and the dynamics 
of gravity play a crucial role, see e.g. \cite{hawk, bd, wald}: What is the 
validity of 
the ``semi-classical'' Einstein's equations
\be
R_{\mu \nu} -  g_{\mu \nu} {R / 2} = 8 \pi G \; \bra{\Psi} \hat T_{\mu \nu} \ket{\Psi}\quad ?
\label{semicl}
\ee
This question concerns the nature  and the
validity of the 
approximations that deliver
eq. (\ref{semicl}) 
starting from the solutions of the Wheeler-DeWitt equation.
There are 
two different (but related) approximations 
involved in this reduction process.
First, the gravitational wave functions must be described 
by WKB solutions so as to have their phases governed 
by the gravitational Hamilton-Jacobi action. 
Secondly, the spread in energy-momentum 
of the 
state $\ket{\Psi}$ 
should be small enough so as 
to legitimate the description
of its evolution
by the Schr\"odinger equation 
\be
i \partial_t \ket{\Psi} = \int \!d^3\!x \; N^\mu \hat {\cal{H}}^m_\mu(g_{ij}) \ket{\Psi} 
\label{schrodi}
\ee
in {\it the} gravitational 
background
solution of eq. (\ref{semicl}).
The 4-D metric $g_{\mu \nu}$
has been split in the
3+1 decomposition\cite{mtw}: $N^\mu$ is the lapse-shift 4-vector and
$g_{ij}$ the metric of the spacelike hypersurfaces.
$\hat {\cal{H}}^m_\mu(g_{ij})= \sqrt{g} \hat T^0_\mu$ 
is the energy-momentum density
operator acting on matter states.

To address the validity of this reduction to a single 
geometry, 
one should in principle 
first construct matrix elements of $\hat T_{\mu \nu}$ in an enlarged
framework in which each matter state is entangled with
its own gravitational wave. Indeed, the Wheeler-DeWitt equation
implies that each matter state determines its own 
gravitational wave (up to the specification of the
quantum state of the
linearized gravitons state that might be considered as 
part of the matter states
in the present discussion).
Only then one can ask in
which circumstances it is legitimate to
replace these entangled waves by a single one valid for all
matter states. 

At this point it should be stressed that this procedure is not
the one which has been generally adopted in the 
literature\cite{HH}-\cite{bk}.
Indeed, in the ``conventional'' treatment, it is {\it a priori} assumed that
the solutions of the WDW 
equation can be factorized into a single WKB 
wave function describing {the} gravitational background
and something which is identified, to the lowest order 
in the inverse Planck mass, as the matter wave function evolving in 
this background according to the Schr\"odinger equation\footnote{
See in particular the recent article \cite{ortiz}: ``In this way
the complete solution of the Wheeler-DeWitt equation is split into
a product of a purely gravitational piece and a mixed piece representing
quantum field theory on that background.''}.
Higher order terms 
determine the
 gravitational corrections to the matter propagation\cite{kiefer, bk}.
The main problem of this approach is that a part
of these 
corrections is of purely classical character and
directly follows from the choice of the  
gravitational wave which has been factorized.
In view of the present difficulties in the interpretation 
of the solutions of the WDW 
equation\cite{vil, isham, wdwin},
one should carefully
determine the origin of the 
 corrections:
Are they intrinsic to the WDW 
equation or are they induced by the 
approximations that have been applied to its solutions?
This question is particularly relevant upon considering
violations of unitarity\cite{fengl}\cite{kiefer}.  

To clarify these issues, another procedure has been 
adopted in \cite{wdwgf, wdwpt, wdwpc}. 
The solutions of the WDW equation have been 
expressed as entangled superpositions of matter and gravitational
waves and the evolution 
of the coefficients of this decomposition has been analyzed
without factorizing a gravitational wave common
to all matter states. {\it After} having obtained 
the dynamical equation governing this evolution, 
a gravitational background
can be introduced 
by performing a first order expansion in the 
matter energy change.
Then, the new expression of the dynamical equation reduces
to the Schr\"odinger equation.
Moreover, in order to minimize the corrections with respect to the 
exact background-free evolution, the background
must be driven by the energy of the matter states under
investigation.
Finally, no violation of unitarity is produced by the 
passage to the new
 description based on a background.

In the present paper, we 
pursue 
this analysis
which was restricted to mini-superspace. 
Our aim is to 
explicitize the specific aspects introduced by
the implementation of
a Background Field Approximation (BFA) to gravity.
In order to separate these aspects from the problems
associated with the quantization of gravity,
we shall work in a classical framework.
However, 
we shall lead 
our investigation
so as to shed light on the quantum problem
which 
consists in obtaining eq. (\ref{semicl}) 
starting from the solutions of the WDW equation.
To this end, we shall analyze
its classical counterpart:
Starting from the Hamilton-Jacobi action
which describes
the entangled evolution of matter and gravity,
what is the nature and the validity of the approximations 
that lead to the matter 
action satisfying the usual Hamilton-Jacobi equation in
a given background whose evolution is fixed?
Specifically,
we shall phrase this question
in path integral terms
by comparing neighboring Hamilton-Jacobi actions
at fixed gravitational end points but differing 
in matter energy content. 
Having chosen this comparison, to apply a BFA
to gravity is a well 
defined mathematical procedure. 

To first order in the change of the 3-geometries
interpolating from the initial to the final condition,
the linear change of the gravitational action equals
the matter energy change in the background geometry.
This is how one recuperates the usual HJ action governing
matter propagation in a given background starting from the
full HJ action, solution of the constraint equations ${\cal{H}}^{gravity}_\mu+
{\cal{H}}^{m}_\mu=0$:
The matter energy term $\int d^3x N^\mu {\cal{H}}^{m}_\mu$ (c.f. eq. (\ref{schrodi})) is 
delivered by the linear ``recoil'' of the gravitational $\int p \dot q dt$ term.
We recall that the full HJ action
contains only 
the sum of the $\int p \dot q dt$ terms for gravity and matter.
This recovery of the matter (light) HJ action is not specific to gravity+matter
systems.
It occurs whenever one applies a BFA to the heavy sector
of an enlarged entangled
system composed of ``heavy'' and  ``light'' degrees of freedom,
see \cite{Bfa}.

This classical result is just what we need in quantum
cosmology. Indeed, the matter part of the full kernel,
solution of the WDW equation, will obey a Schr\"odinger equation
when the gravitational waves are WKB 
and when
a first order expansion in the gravitational change makes sense.

The validity of this first order expansion is controlled
by 
higher order terms.
These determine the gravitational corrections to the usual 
HJ matter action and their importance establishes {\it a posteriori} whether
it was legitimate to treat gravitational degrees of freedom as ``heavy''. 
They may 
be analyzed in classical settings. 
This crucial point deserves more explanations: The second order term
has a structure of the form $\Delta \e^2 \times \partial_\e^2 S_{G}$
where $S_{G}$ is the gravitational HJ action
evaluated on the background. However, in quantum cosmology,
the origin of the spread in energy $\Delta \e$ may be purely quantum.
Therefore, the structure of these corrections is fully determined
by the classical analysis whereas its normalization may not be. 
This explains why we shall ``recover'' expressions 
that have been
obtained
in the quantum analysis\cite{fengl, bk}.
Moreover we shall show that the corrections to the 
WKB gravitational wave functions are governed by
expressions similar to those governing the BFA.
This legitimizes the implementation of a BFA to WKB
variables.  

Finally, the analysis of this classical problem has
its own interest. It makes bridges between the techniques used 
when studying growth of local fluctuations (in particular as phrased by 
Steward and Salopek\cite{ss}) and those used
in quantum cosmology.
Moreover it also sheds light on the 
``issue of time'' in (quantum) cosmology\cite{isham,
bprmp}.
Our analysis indeed does not coincides 
with the similar investigation performed by Barbour\cite{barbour}.

In this article, we proceed in three steps.
In Section 2, we analyze the action in the
simple case of an homogeneous matter field
in minisuperspace. Furthermore, we restrict ourselves
to matter hamiltonians such that the amplitude of the 
field stays constant. 
Then the implementation of the 
BFA can explicitly and 
easily be performed in gauge independent terms.

In Section 3, we work in a minisuperspace model
in which the background is characterized by many degrees of freedom
and we no longer put restrictions on the matter hamiltonian.
Moreover we make use of techniques very similar
to those required by the analysis of the solutions of 
the WDW equation.  
 
In Section 4, we 
generalize these results to arbitrary 3-geometries. 
We shall also comment on the treatment of second order
corrections to the BFA which
 has been
recently presented in \cite{bk}.


\section{The background field approximation when 
matter is characterized by a constant of motion}\label{ad}

As a warming up, we first show
how a background field contribution is chosen and extracted from 
the HJ action describing the entangled evolution
of matter and gravity. 
Then, we show how this choice
determines the nature of the corrections
to the description of {\it matter} evolution.
To explicitize these corrections, 
we begin this Section by the analysis of the HJ action of 
matter and gravity without making any approximation,
i.e. in a background free description.
In order to simplify this analysis,  
we shall limit ourselves to cases in
which the matter 
energy is determined by a constant of motion.

We recall that in this Section and the next one, we work
in minisuperspace. For these geometries, 
the metric element 
can be written as 
\be
ds^2 = - N^2(\xi) d\xi^2 + a^2(\xi) d^2 \Omega_3
\label{M2}
\ee
where $N(\xi)$ is the lapse function, $a(\xi)$ the scale factor  and
$d^2 \Omega_3$ the constant line element of the
homogeneous three surfaces.

The simplest matter system whose evolution
is characterized by a constant
of motion consists of an homogeneous distribution 
of comoving oscillators.
Classically, these oscillators can be
represented be the zero-component momentum of
a massive scalar field\cite{wdwgf}.
Their action is
\ba
S_M &=&  \int d\xi { N(\xi) }  \left( 
{(\partial_\xi \phi)^2   \over 2 N^2(\xi) }- { M^2  \phi^2 \over 2} \right) 
\nonumber\\
&=&
 \int d\xi  \left\{ p_\phi \partial_\xi \phi
- { N(\xi)} \left( p_\phi^2  /2 + M ^2  \phi^2 /2 \right)  \right\} 
\label{M4}
\ea
where $M$ is the frequence of the oscillations.
On the equation of motion, 
their energy with respect to $\xi$ is 
\be
N(\xi) \e = N(\xi) { M }  |A_M |^2 
\label{M5}
\ee
where $A_M$ is the constant amplitude of the field.

Another example consists of
conformally coupled scalar photons of fixed (conformal)
momentum $k$. 
On the equation of motion, their energy is
$N \e_\gamma = N k |A_\gamma |^2 /a$, see \cite{wdwgf}.
These examples can be generalized by 
considering non quadratic hamiltonians in the field amplitude
or by considering oscillations with an arbitrary $a$-dependent
frequency $\om(a)$ treated in the adiabatic approximation 
so as to 
guarantee that their 
energy is $N(\xi) \e_m(a) = N(\xi) { \om(a) }  |A_m |^2$.
One can of course also consider a sum of these systems.

\vskip 2. truecm
{\bf The background-free description}
\vskip .1 truecm
\noindent
The residual Bianchi indentity
(see Chap. 27 in \cite{mtw})
requires that matter satisfies
\be
\partial_a \e_m(a) = - 4 \pi P(a) a^2
\label{bianch}
\ee
where $P$ is the pressure.
In all the above examples, this equation is satisfied because
$\e_m(a)$ characterizes on-shell matter propagation.
Of crucial importance is the fact that the matter energy
$\e_m(a(\xi))$ depends on $\xi$ through $a$ only,
thereby allowing eq. (\ref{bianch}) to have this intrinsic
lapse-free writing.

In minisuperspace, the gravitational hamiltonian 
is given by 
\be
N(\xi) H_G = N(\xi) { 1 \over 2aG} \left(
-G^2  \pi_a^2  + \kappa a^2 + \Lambda a^4
\right)
\label{hg}
\ee
where $G$ is Newton's constant, $\pi_a$ the momentum
of $a$, $\kappa$ is equal to $0, \pm$ for 
flat, open or closed three surfaces
and $\Lambda$ is the cosmological constant. 
Thus when the matter energy $\e_m(a)$ is given, 
the HJ constraint equation, $H_G + H_m = 0$, reads 
\be
-G^2  ( \partial_a S_G (a, \e_m  ) )^2
 + \kappa a^2 + \Lambda a^4 + 2 Ga \;
\e_m (a)  = 0
\label{M12}
\ee
The lapse $N$ has also disappeared from this equation.
In this Section, it will not be re-introduced thereby guaranteeing a gauge
independent description.

The HJ action $S_G$, 
solution of eq. (\ref{M12}), is 
\be
S_G(a_2 , \e_m  ) = 
\int_{a_1}^{a_2} da \; \pi( a, \e_m  )
\label{M8}
\ee
where $\pi( a, \e_m)$ is the on-shell
momentum driven by $\e_m(a)$
\be
\pi ( a, \e_m  ) = - G^{-1} \sqrt{ 
\kappa a^2 + \Lambda a^4 + 2 Ga \e_m (a)}
\label{M9}
\ee
Notice that the minus sign 
characterizes
expanding universes. It arises from the unusual
sign of the kinetic gravitational energy in eq. (\ref{M12}).

The total HJ action (gravity + matter), solution of the
constraint equation, is thus
\be
S_T = S_{G}(a_2, \e_m) + S_{m}(\phi_2, \e_m)  = \int_{a_1}^{a_2} da \; \pi( a, \e_m  ) + 
\int_{\phi_1}^{\phi_2} d\phi \; p_\phi( \phi , \e_m  )
\label{tots}
\ee
where $p_\phi( \phi , \e_m  )$ is the matter momentum,
determined by the energy condition $H_m(p_\phi, \phi, a) = \e_m (a)$.
For definiteness and simplicity, from now on 
 we consider only the case of
harmonic oscillators.
In this case, one has
$p_\phi( \phi , \e  )= \sqrt{2 \e  - M^2 \phi^2}$,
see eqs. (\ref{M4}, \ref{M5}) and compare it with eq. (\ref{M9}).
Conformal photons or more general oscillators 
are treated along similar lines.

At this point we have imposed that the total energy
vanishes and that both gravitational and matter 
propagations occur along classical orbits.
However, when working with fixed the end point conditions 
$(a_1, \phi_1; a_2, \phi_2)$, 
$S_T$
is not fully extremal:
It should still be extremized with respect to 
variations of $\e$\cite{bprmp, lanc}. 
This stationary condition determines
the ``saddle point'' value $\bar \e=  \bar \e(a_1, \phi_1; a_2, \phi_2)$
and reads
\ba
-\partial_{\e} \int_{a_1}^{a_2} da \; \pi( a, \e) \vert_{\e = \bar \e }
&=& \partial_{\e} \int_{\phi_1}^{\phi_2} d\phi  \; p_\phi( \phi , \e ) \vert_{\e = \bar \e }
\nonumber\\
- \int_{a_1}^{a_2} da { a \over G \pi( a, \bar \e) } 
&=& \int_{\phi_1}^{\phi_2} d\phi { 1 \over p_\phi( \phi ,  \bar \e )}
\label{bul}\\
t(a_2, \bar \e) &=& t_{cesium}(\phi_2, \bar \e) 
\nonumber 
\ea
On the left hand side, the gravitational integral has defined $t(a_2, \bar \e)$,
the 
proper time lapse from $a_1$ calculated
from the propagation of $a$ driven by the ``saddle point''
energy $\bar \e$.

On the r.h.s.
the matter integral has defined the 
``cesium'' time $t_{cesium}(\phi_2, \bar \e)$. In the 
case of massive harmonic oscillators
at rest, it equals to the period $2 \pi /M$ times
the number\footnote{There might be a discrete 
number of solutions to eq. (\ref{bul})
since we are dealing with periodic matter systems. 
However this multiplicity does not affect what follows
 since we can always choose to work with the lowest value.}
of periods. 
(In the case of conformal photons, one
would have obtained respectively 
the 
conformal time on the l.h.s. 
and the number of conformal periods on the r.h.s., 
c.f. \cite{wdwgf}.)

Is it through the equality of these
times that one obtains the {orbit} 
$\phi_2= \phi(a_2, \bar \e)$
characterized by $\bar \e$ and which 
passes by $\phi_1$ at $a_1$.
Once this correlation is obtained, it can be recast 
in the usual way parametrized by $t$: 
$\phi_{\bar \e}(t), a_{\bar \e}(t)$. In this writing,
the dependence in $\bar \e$ is ``blamed'' on the dynamical
variables. 

The unusual aspect in this derivation
of the trajectory is that 
$t$ has arisen {\it a posteriori} 
through the dynamics of both matter and gravity. 
This is because the usual linear term in $E t_{external}$ 
was not present in the total action $S_T$, eq. (\ref{tots}).
Rather both $S_G$ and $S_m$ depend non-linearly on $\e$.
Up to now indeed $S_G$ and $S_m$ have been treated on the same
footings. By the implementation of a BFA,
 this symmetry will be broken since the 
BFA will concern only $S_G$. 
In other words,
only $a$ will be treated as an ``heavy'' degree of freedom.
Then, the notion of an ``external'' time to matter (light) dynamics 
will be recovered from the heavy character of $a$.
Notice that the division into light and heavy
variables is not a question of taste: The physical
circumstances decide whether it makes 
mathematical sense.
Notice also that this procedure based on a division 
between light and heavy variables 
does not coincide 
with Barbour's approach\cite{barbour} to time in cosmology.
In his approach, time is a redundant variables
which depends on all degrees of freedom. 

\vskip 1.3 truecm

{\bf The background field approximation}
\vskip .1 truecm
\noindent
We now apply a BFA to the gravitational part of
the full HJ action $S_{G}+ S_{m}$.
To this end,
one should first choose the
reference background.
To choose it, one needs to focus on 
a given ``physical'' problem.
Indeed the choice of the background should be tided up to
the problem under investigation and made
so as to minimize the errors
induced by the approximation process. 
This procedure has been applied in quantum 
cosmology\cite{wdwpt}:
Upon computing
a given transition amplitude from a initial matter state
to a final one, it was shown that the ``best'' background is
determined symmetrically by the energy contents
of the initial and final states. Any other choice leads 
to greater systematic deviations with respect to the exact 
(background free) description.

We use the same philosophy in the classical framework:
We want to determine the simplified description of 
matter propagation in a {\it single} background geometry
when the final values of $\phi_2$
are centered around a given $\phi^B_2$ and when the 
three other conditions ($a_1, \phi_1; a_2$) are held fixed.
This imposes to work with the background 
determined by the end point values 
($a_1, \phi_1; a_2, \phi^B_2$)
since any other choice leads 
to systematic deviations.
To obtain the simplified matter description,
 we expand only the gravitational part of the HJ action
in power of $\Delta \e= \e - \bar \e^B= \e -
\bar \e( a_1, \phi_1; a_2, \phi^B_2)$.
(From now on, we use the subscript $B$ to identify 
quantities evaluated on their background value.)
Notice that we have kept $\e$ free 
since 
the new description of matter evolution still
involves the extremization of $S_T$ with respect to it. 

By developing $S_G$ to first order in $\Delta \e= \e - \bar \e^B$, one obtains
\be
S_G(a_2, \e)
+ S_M(\phi_2, \e) = 
S_G(a_2, \bar \e^B)
+  \int_{a_1}^{a_2} da \; (\e - \bar \e^B) \partial_{\e} \pi( a, \e)\vert_{\e = \bar \e^B} 
+ S_M(\phi_2, \e)
\label{fo2}
\ee
As in the absence of approximation,
$S_{G}+ S_{m}$ must be 
stationary
with respect to changes in $\e$.
Then, as before, $\partial_\e S_T = 0$ determines the 
saddle point energy $\bar \e$.
However, the new expression of this condition now reads
\be
t(a_2, \bar \e^B) =  \int_{\phi_1}^{\phi_2} d\phi { 1 \over p_\phi( \phi , \bar \e )}
\label{bul2}
\ee
in the place of eq. (\ref{bul}).
The {\it only} difference with that equation is that 
$\bar \e$ on the l.h.s. has been replaced by $\bar \e^B$ 
for all values of $\phi_2$.
Thus $ t(a_2, \bar \e^B)$, the background time, 
is external to matter dynamics and the orbit directly
expressed as $\phi_2 = \phi_{\bar \e} (t)$.

The errors induced by this BFA are of two kinds:
First $\bar \e$, solution of eq. (\ref{bul2}), differs 
(slightly when $a$ is heavy) from the solution of eq. (\ref{bul})
and secondly, $t(a_2, \bar \e^B)$ is not the time in
the actual geometry. This is the price to pay when one has
chosen to 
work with a given background common for different values of $\phi_2$.
It should be noticed that neither error affects
the orbit $\phi_2 = \phi_{\bar \e} (t)$ since both are 
concerned with the choice of the orbit through 
$\bar \e$ and the parametrization
of $a_2$ by $t(a_2, \bar \e^B)$. 
This absence of deformations of the orbit
follows from the existence of the constant $\e$ which allows to
separate the action $S_T$ into two unconnected pieces. 
This will no longer be the case in the next Sections.

Eq. (\ref{bul2}) has been obtained from the HJ action, eq. (\ref{tots}),
 in three steps.
One has first chosen the reference background driven by $\bar \e^B$.
One has then expand the gravitational part of the action
to first order in $\e- \bar\e^B$ and finally requires 
extremization with respect to $\e$.
This is how the
stationary condition\footnote{
In the language of MTW\cite{mtw}, it is called the 
``condition of constructive
interference'' in reference to the machinery of
quantum mechanics. Upon working
in quantum cosmology,
we shall see indeed that both eqs. (\ref{bul}, \ref{bul2})
appear with this status.}
determining
matter propagation in {the} background $a^B(t)= a_{\bar \e^B}(t)$ 
is obtained from the total action $S_G + S_M$
governing the entangled dynamics of 
matter and gravity.

This procedure 
generalizes the usual test-particle (or test-field) approximation 
 for two reasons. 
First, in that approximation,
there is no influence of the matter energy on the 
determination of the background evolution. Here instead,
the ``mean'' matter energy $\e^B$ does determine
$a^B(t)$, see eq. (\ref{M9}).
Secondly, because of this partial
backreaction effect, the corrections
to the BFA differ from those to the test-particle approximation,
see below.



\vskip .3 truecm

{\bf The validity of the background field approximation}

\noindent
The validity of the approximate description based on eq. (\ref{bul2}) 
is governed by the {\it influence} of the higher order terms in $\Delta \e$.
The quadratic term of the expansion of
$S_G(a_2, \e )$ 
around the background contribution 
$S_G(a_2, \bar \e^B)$ is 
\ba
{ (\Delta \e)^2 \over 2}\partial^2_\e  S_G(a_2, \e)\vert_{\e = \bar \e^B} 
&=&
{ (\Delta \e)^2\over 2} \partial_\e t(a_2, \e )\vert_{\e = \bar \e^B} 
= {(\Delta \e)^2 \over 2}\int_{a_1}^{a_2} {da  a
\over G}
{\partial_\e \pi(a, \e) \over
\pi^2(a, \e) }\vert_{\e = \bar \e^B} 
\nonumber\\
&=&
\int_0^{t^B(a_2)} dt (\Delta \e) \left( { 1 \over 2} { G \Delta \e
\over \kappa a^B(t) + \Lambda [a^B(t)]^3 + 2 G \bar \e^B} 
\right)
\label{quadr}
\ea
The second expression makes clear that this correction
is governed by the ``susceptibility'' of the background geometry.
It is indeed 
controlled by the amount of change in the background time lapse
$dt = d\e \partial_\e t$ induced by the
change in the matter energy $d\e$.

In the last expression, we have factorized the contribution of
the linear term ($dt \Delta \e$) so as to identify the 
relative correction to it. 
In order to give an estimate to this correction factor,
it is appropriate to consider matter dominated universes
 for which $G \bar \e^B >\!\!> 
\kappa a + \Lambda a^3$. Then the correction
factor is simply
$\Delta \e /4\bar \e^B$: the change in the matter energy
divided by the matter energy $\bar \e^B$ which drives the 
background. 
Notice that this ratio is independent of $G$.
Thus
the BFA is not an expansion in powers of $G$.
Similarly, in quantum cosmology, we shall see
that the BFA is not an expansion in the inverse Planck mass.
On the contrary, the test-particle approximation 
can be viewed as the linearized theory in $(G \e)$
since $\e$ does not enter into the determination
of the background.
Eq. (\ref{quadr}) also shows that the best background 
geometry to describe matter propagation when 
the values of $\phi_2$ are centered around $\phi_2^B$ 
is determined by $\bar \e^B=\bar \e(a_2 , \phi^B_2)$.

To determine the influence of the quadratic term, 
one should add it
to the l.h.s. of eq. (\ref{bul2}) and compute the 
new value of the saddle energy $\bar \e$.
One finds that its effect
is to shift the 
saddle point energy $\bar \e$ by $(\bar \e - \bar\e^B) \partial_\e t^B
/\partial_\e t_{cesium}$.
Thus, when $\phi$ is a light and rapidly changing 
variable when compared to $a$, this shift is 
negligible since
$\partial_\e t^B <\!\!<
\partial_\e t_{cesium}$. This inequality shows that
the value of the correction term
does not control alone the validity of the BFA for describing
matter evolution.
Indeed
this validity is governed by a ratio in which  
the lightness of $\phi$ also intervenes. 

The fact that the change of $\bar \e$
is controlled by a ratio of ``specific heats''
is reminiscent to statistical mechanics:
Consider a microcanonical ensemble of energy $E$
 composed of two systems $1$ and $2$ with densities of states 
given by 
 $\Omega(E_1)= e^{S(E_1)}$ and $\om(E_2)=e^{s(E_2)}$. 
When evaluating the density of states of the
total system given by $\int d\e \Omega(E-\e) \om(\e)$
by a saddle point approximation, one obtains the
equality of the temperatures: $\partial_\e S(E - \e) + \partial_\e s(\e) =0$
which is the equivalent of eq. (\ref{bul}), see App. B in \cite{wdwgf} 
for more details.
Moreover, close to equilibrium and when $E-\bar \e \gg \bar \e$, one can
treat the big system $1$ at the BFA, i.e. develop $S(E)$ to first order
in $\Delta \e$. Then the errors induced by this canonical
approximation are governed by the ratio of the 
specific heats $\partial^2_\e S/ \partial^2_\e s$. For homogeneous
systems this ratio scales like the ratio of the volumes.
Therefore the BFA can be considered as a large reservoir limit.

We conclude this subsection by mentioning
 an interesting question\cite{turnip}:
What happens close to a turning point?
In this case the quadratic correction term, eq. (\ref{quadr}),
becomes unbounded since the gravitational 
momentum $\pi(a, \bar \e^B)$
vanishes.
This signals the ``instability'' of 
the parametrization of matter evolution by $t(a, \bar \e^B)$.
Thus, close to a turning point,
$t(a, \bar \e^B)$ is not a good 
 parameter.
A very good parameter is $t(\pi_a, \bar \e^B)$ 
obtained by performing a Legendre
transformation to $S_G$ so as to fix the end value of the momentum
rather than $a_2$. A similar situation occurs upon considering the
validity of the WKB approximation: 
Close to a turning point,
this approximation becomes also completely 
unappropriate. However the
physics is not directly affected since, it is the state (the ket)
and not its expression in the position representation 
which governs physical processes through matrix elements.


\vskip .3 truecm

{\bf Application to quantum cosmology}
\vskip .1 truecm
\noindent
The application of these results to quantum cosmology
is straightforward if one postulates that gravitational waves 
are WKB. In this case indeed, their phases and norms are determined
by the gravitational HJ action, eq. (\ref{M8}).
At first sight it might seem that the validity of the
WKB approximation introduce new restrictions.
However, as pointed out in \cite{vil} and further
clarified in \cite{wdwgf}, the conditions legitimizing the use 
of WKB gravitational waves are 
concomitant with those 
governing the validity of the BFA:
Both approximations are valid
when the relative change of the 
kinetical energy of the scale factor is negligible.

Upon quantizing matter systems governed classically
by constants of motion, one obtains stationary eigenstates
labeled by a discrete parameter $n$ ($e.g.$ the occupation number)
and characterized by an eigenvalue:
\be
\hat H_m(a) \ket{n} = \e(n)  \ket{n} 
\label{ean}
\ee
In the time-free description, stationarity means
that $\partial_a \ket{n} = 0 $. 
This implies that the kernel, solution of the WDW equation
\be
(\hat H_G + \hat H_m) {\cal{K}}(a, \phi; a_1, \phi_1) = 0
\label{hatt}
\ee
can be decomposed as
\be
{\cal{K}}(a, \phi; a_1, \phi_1) = \sum_{n} {\cal{K}}_{n} (a; a_1) \; K_n ( \phi; \phi_1) 
\label{ker}
\ee
wherein the matter kernel $K_n ( \phi; \phi_1)$
 is equal to $\scal{\phi}{n}\scal{n}{\phi_1}$
as in Schr\"odinger settings.
Moreover since the structure of ${\cal{K}}$ is still
a sum of products, 
 the BFA will concern only the gravitational 
kernels  ${\cal{K}}_{n} (a ; a_1)$.
This is true only for
 the simple mini-superspace cases we are dealing with.

The gravitational kernel ${\cal{K}}_{n} (a ; a_1)$ is a solution
the quantized version of eq. (\ref{M12}), i.e.
\be
\left(
G^2 \partial_a^2
 + \kappa a^2 + \Lambda a^4 + 2 Ga \;
\e(n)\right)   {\cal{K}}_{n} (a ; a_1)  = 0
\label{M12'}
\ee
In the WKB approximation, for $a>a_1$, it is equal to
\be
{\cal{K}}_{n}(a; a_1) = { \exp \left[ i \int_{a_1}^{a} da' \; \pi( a', \e(n) ) \right]
\over
\sqrt { \pi( a , \e(n)) \pi( a_1, \e(n))}}
\label{kerkb}
\ee

We now determine the behavior of 
kernel ${\cal{K}}(a, \phi; a_1, \phi_1)$
when $a$ is equal to $a_2$ and $\phi$ close to 
$\phi^B_2$.
To make contact with 
 classical mechanics,
we use a saddle point approximation
(i.e. a constructive interference condition)
to estimate 
the summation over $n$
in eq. (\ref{ker}). 
When using the WKB expressions eq. (\ref{kerkb}), 
the equation which fixes the saddle point value $\bar n^B$
is equal to that one would have obtained by treating
gravity at BFA if the background was the
solution driven by 
the saddle point energy $\e^B =\e(\bar n^B)$.
This is clearly exhibited by 
factorizing ${\cal{K}}_{\bar n^B}(a_2; a_1)$ and developing
the gravitational
kernels in $\Delta n= n - \bar n^B$
\ba
{\cal{K}}(a_2, \phi^B_2; a_1, \phi_1)
&=& {\cal{K}}_{\bar n^B}(a_2; a_1) \times \sum_n
K_{n} ( \phi^B_2 ; \phi_1)
\sqrt{ {\pi( a_2 , \e(\bar  n^B)) \pi( a_1, \e(\bar n^B))
\over \pi( a_2 , \e(n)) \pi( a_1, \e(n)) }}
\nonumber\\
&&
\exp \left[ i \int_{a_1}^{a_2} da \; \left\{ \pi( a, \e(n))- \pi( a, \e(\bar n^B) ) 
\right\}\right]
\nonumber\\
&=&
{\cal{K}}_{\bar n^B}(a_2; a_1) \times \sum_n
K_{n} ( \phi^B_2 ; \phi_1)
 \left\{  1 + O \left( {\Delta_n \e \over \pi^2 (a, \e(\bar n^B))}\right) 
\right\}
\nonumber\\
&&
\exp \left[ -i \int_{0}^{t_{\bar n^B}(a_2)} dt \left\{ \e(n)- \e(\bar n) \right\}
 \left\{  1 + O \left( {\Delta \e_n \over \pi^2}
\right) \right\} \right]
\label{schroo}
\ea
where $\Delta_n \e = \e(n) - \e(\bar n^B)$.
The stationary phase conditions are respectively
\ba
\partial_n \left\{ \int_{a_1}^{a_2}\!da \pi( a, \e(n)) + i \ln K_n (\phi^B_2; \phi_1)
\right\} &=& 0
\label{spc20}
\\ 
\left({d \e(n) \over dn} \right)
t_{\bar n^B}(a_2)
 + i 
\partial_n \ln K_n ( \phi^B_2; \phi_1)
 &=& 0
\label{spc2}
\ea
Eq. (\ref{spc20}) is obtained from the first expression
of eq. (\ref{schroo}) and does no rely on any BFA.
It can be viewed as the quantum version of 
eq. (\ref{bul}). 
Instead eq. (\ref{spc2})
follows
from the second expression
in which a BFA has been already performed, {\it c.f.} eq. (\ref{bul2}).
This later condition is the usual ``constructive interference condition''
obtained from QFT in the background characterized by 
$t_{\bar n^B}(a)$.
By construction, these conditions coincide 
since the expansion in $\Delta n$ was performed 
precisely around the saddle point value $\bar n^B$.

From this agreement and the former classical analysis
leading to eq. (\ref{bul2}), one immediately deduces
that the saddle point phase of the kernel
considered as a function of $\phi_2$ in a neighborhood of
$\phi_2^B$ is equal to the BFA phase
when non linear terms in $\bar n - \bar n^B$ are discarded.
(Notice that these correction terms
are governed by $\Delta_n \e/ \pi^2$ as in the classical case.)
Moreover, when they are discarded,
the {\it spread} in $n - \bar n$ agrees with the BFA 
spread since in both cases (with or without BFA)
it is determined by the (second derivative of the log of the)
 matter kernel $K_n ( \phi; \phi_1)$ only.
Notice that this spread is {\it intrinsic}, i.e.
 determined by the physical circumstances
and $\hbar$. This is to be opposed to the classical
spread $\Delta \e$ in the former subsection which 
was fixed by the ``external'' choice of $\phi_2$.
However, the mathematics involved are the same.

The above three agreements imply that 
${\cal{K}}(a_2, \phi_2; a_1, \phi_1)$
satisfies the Schr\"odinger equation in the background geometry
$a^B(t)$
for $\phi_2 $ 
sufficiently close to the background
value $\phi^B_2$ so as to legitimize the neglection 
of higher order terms in $\Delta n$.
Notice however  
the unconventional property of this Schr\"odinger equation.
Since the background geometry is dynamically determined by $\bar n$,
c.f. eq. (\ref{spc20}), it changes when $a_2$ varies
except if one tunes $\phi_2^B$ so as to stay on the {orbit}
$\phi_2 = \phi( a_2, \bar n^B)$.
Because of this, it is appropriate to work with
fixed $n$ rather than with fixed end point values on the matter field. 
This is what has been adopted in \cite{wdwpt, wdwpc}
to study transition amplitudes in quantum cosmology.
This is also what we shall adopt in the next Sections
in the definition of the background.

As in classical mechanics, higher order terms in $\Delta n$ (or $\Delta \e$)
induces corrections to the determination
of the saddle point value $\bar n$. 
To determine their physical consequences it is mandatory
to introduce interactions leading to transitions among
states, see \cite{wdwpt, wdwpc}.
The main lesson of these works is that it is not necessary
to first factorize the mean gravitational kernel 
and then search for the description of matter evolution\footnote{
It is even a nuisance!
The reason is that transition amplitudes are independent 
of the mean energy: they only depend of the energy of the
two states characterizing the transition. However, when searching
for the description of these transitions in the background geometry
driven by the mean energy, one obtains series in which all 
terms depend parametrically on this energy. This is nothing but the 
consequence of having first factorized the mean gravitational kernel.
By no means it is an intrinsic properties of the solutions
of the WDW equation.}.
Indeed, the WDW equation determines their evolution 
directly in terms of $a$, i.e. in a background free description, 
see eq. (\ref{ker}) for a simple example. 
However, as in eq. (\ref{schroo}),
one can always {\it a posteriori}
perform a BFA (an expansion in $n- n^B$)
in order to obtain a more conventional
description formulated in a single background\cite{bprmp}. 
Doing so one induces non linear terms in $n- n^B$ whose role
among other things
is to introduce phase shifts governed by $\partial^2_\e S_G$.

In resume, eq. (\ref{schroo}) shows that the corrections 
to the kernel with respect to its Schr\"odinger BFA expression
are of two kinds: those due to quadratic and 
higher order terms in $\Delta n$
and those due to the WKB approximation, eq. (\ref{kerkb}). 
In a next paper, we shall analyze the consequences
of both types of corrections on possible violations of 
unitarity.


\section{The background field approximation
for general mini-superspace models}

In this Section we no longer impose that matter
dynamics are governed by a constant of motion.
Therefore we are
obliged to apply the BFA to extremized actions since we have lost 
the possibility of applying a BFA before
having extremized the total action
with respect to $\e$, see eq. (\ref{fo2}). 
There are two ways to implement this approximation. 

The first one consists in keeping the lapse function $N(\xi)$ and working
with trajectories parametrized by $\xi$. The main advantage 
of this approach is that the whole space-time structure
of the background emerges upon applying the BFA to gravity.
Indeed since both $a^B(\xi)$ and $N^B(\xi)$ are determined, one
obtains matter propagation in a curved background
which possesses a given time parametrization.
A second advantage is that it offers the 
possibility of considering
``off-shell'' matter configurations.
At first sight it might seem completely
absurd to consider the
gravitational reaction to off-shell matter configurations
since r.h.s of Einstein's equation will be incompatible with the Bianchi 
identity, i.e. eq. (\ref{bianch}) will not be satisfied.  
However, in the path integral,
most of the matter configurations from $\phi_1$ to $\phi_2$ do
violate this equation. Therefore it is a legitimate question
to ask whether the action governing these configurations
can be conceived as arising 
from an enlarged action governing the
evolution of both gravity and matter.
This analysis is performed in Appendix B.

The second approach is more intrinsic since it is based
on the HJ action rather than on trajectories. It does not 
make use
of the lapse function and furthermore 
proceeds along lines very similar to those used in
quantum cosmology. Therefore, we shall present here
this second approach.
The first one is developed in Appendix A.
The reader unfamiliar with 
HJ techniques is invited to first read this Appendix
which procceds along more conventional
mechanical lines.

In order to prepare the analysis of the next Section
which deals with general 3-surfaces, we 
work with backgrounds described by many degrees of freedom, $X^i$.
These consist on the scale factor $a$, on the homogeneous
part of some massive inflaton field and possibly also
on gravitational Fourrier modes\cite{HH} highly excited.
The multidimensional character of the background trajectories
will introduce new difficulties since there are now 
``transversal'' directions to the background orbits\footnote
{It might seem a little bit antinomic to consider a plurality 
of backgrounds. Moreover, there is a simple
way to reject this plurality:
It consists in treating all but one coordinates $X^i$ (say e.g. $a$)
as ``matter'' degrees of freedom. One would then recover a 
situation similar to that of the former Section in which the ``matter''
energy univocally specifies the orbits of the remaining 
background coordinate.  
This is effectively what has been adopted by Halliwel and Hawking in \cite{HH}.
In what follows however, we shall
proceed with background orbits characterized by many 
coordinates $X^i$ in order to confront the new 
difficulties.
}\cite{kiefer, ortiz, bk}.

The total hamiltonian we are considering is of the form
\be
N(\xi) H_T= 
N(\xi) \left[ H_X(\pi_i, X^i) + H_m(\phi, p_\phi, X^i )
\right]
\label{TA0}
\ee
We only impose that $H_X$ and $H_m$
be quadratic in the momenta $\pi_i$ and $p_\phi$.
Notice also that we have not included in $H_m$
a ``non-minimal'' coupling to $\pi_i$.
The inclusion of this
additional coupling  presents no difficulty however.

The total action satisfies the HJ constraint equation
\be
{ 1 \over 2}
C^{ij}(X^k) \partial_{X^i} S_T \; \partial_{X^j} S_T + V(X^k)
+ H_m(\phi, \partial_\phi S_T, X^i ) =0
\label{hjt}
\ee
As in eq. (\ref{tots}), $S_T$ can be written as a sum of terms
\be
S_T = S_X + S_m = 
 \int^{X^i_2}_{X^i_1} 
dX^i \pi_i(X^i, \left\{X^j_2, \phi_2\right\})
+ \int^{\phi_2}_{\phi_1}
 d\phi p_\phi(\phi, \left\{X^j_2, \phi_2\right\})
\label{hjact}
\ee
where each on-shell momentum is parametrized by 
its conjugate position. In brackets we have indicated that
they depend parametrically of the end point values of all 
variables. It is on this dependence that the BFA will act.

Before proceeding to the implementation of the BFA,
we should carefully determine the background(s).
In the former Section, this was easily achieved by the
specification of the constant of motion $\bar \e^B$.
As we discussed in the ``quantum cosmology''
subsection, this amounted to work with 
the end value of $\phi$ evaluated
along the orbit characterized by $\bar \e^B$.
In the present situation without constant of motion,
we define the background by fixing 
the initial value of the matter energy
$H_m( a_1, \phi_1, p_\phi)$, i.e. we fix the initial momentum 
$p_{\phi, 1}
= p_\phi^B$. Then the end value of the $\phi$ field
is evaluated along the orbit specified by $p_\phi^B$
as well as the end-point values $X_2^i$. 
Indeed the specification of the matter energy no longer suffices to 
specify univocally the orbit. Therefore the set of $X_2^i$
will both determine the time lapse along the trajectory 
(as $a$ did in the former Section)
but also determine the orbit, i.e. the location in the ``transversal'' 
directions.

With these aspects in mind, we now search for the approximate
description of matter evolution when the {\it change}
in $\delta \pi_i= \pi_i(X^i, \left\{X^j_2, \phi_2\right\})
- \pi_i(X^i, \left\{X^j_2, p_\phi^B\right\})$
induced by the change in the matter configurations
is treated to first order.
To this end, we write 
\be
S_T(X^i_2, \phi_2) = S_X^B(X^i_2) + \Delta S(X^i_2, \phi_2)
\label{dst}
\ee
where $S_X^B(X^i_2)$ is the background action given by
$ \int^{X^i_2}_{X^i_1} 
dX^i \pi_i(X^i, \left\{X^j_2, p_\phi^B\right\})$.
By inserting eq. (\ref{dst}) into the constraint eq. (\ref{hjt})
one obtains
\ba
&&C^{ij}(X^k) \partial_{X^i} S_X^B \; \partial_{X^j} \Delta S
+ H_m(\phi, \partial_\phi \Delta S, X^i ) - H_m^B(X^i ) 
\nonumber\\
&&\quad
+  {1 \over 2}
C^{ij}(X^k) \partial_{X^i} \Delta S\; \partial_{X^j} \Delta S
=0
\label{hjt2}
\ea
where we have used the fact that $S_X^B$ is the 
solution of eq. (\ref{hjt}) driven by the background matter energy
$ H_m^B(X^i )$. Notice that $H_m^B(X^i )$ is 
orbit dependent but nevertheless acts as a potential energy
along each orbit.
Moreover, along each orbit, the proper 
time lapse to go from $X^i_1, \phi_1,
p_\phi^B$ to $X^i_2$ satisfies (by definition)  
\be
C^{ij}(X^k) \partial_{X^i} S_X^B \; \partial_{X^j} t^B(X^i) = 1
\label{prtl}
\ee
This is just another way to express the relationship 
between velocities and momenta which is usually 
written as $dX^{iB}/dt = C^{ij} \pi^B_i$ when one has $t$
at our disposal from the outset. 
Here instead, $t^B(X^i)$ is re-introduced and defined 
by $-\partial_E S^B_X$ where $dE$ is
an infinitesimal constant change of the matter energy $H_m^B$.
Then by varying eq. (\ref{hjt}) applied to $S^B_X$
with respect to $d H_m^B=dE$,
one obtains eq. (\ref{prtl}).

In the present formalism, to apply a BFA consists in neglecting 
the last term of eq. (\ref{hjt2}).
This amounts to assume that 
$\partial_{X^i} \Delta S \ll \partial_{X^i} S_X^B$,
i.e. $\delta \pi_i \ll \pi^B_i$.
Then, using eq. (\ref{prtl}),
the first line of eq. (\ref{hjt2}) reads
\be
 \partial_t \Delta S(\phi, X^{iB}(t))
+ H_m(\phi, \partial_\phi \Delta S, X^{iB}(t) ) - H_m^B(t ) =0
\label{hjt3}
\ee
Thus, to first order in $\delta \pi_i$,
$\Delta S$ satisfies
the HJ equation
for the matter field in the background 
determined by $X^i_2, p_\phi^B$. 
Then, as usual, 
it can be decomposed as
\be
\Delta S(\phi_2, t) = \int^{\phi_2}_{\phi_1}
 d\phi p_\phi(\phi, \left\{
\phi_2\right\}) - \int^{t}_0
dt' \left\{H_m(\phi, p_\phi, X^{iB}(t') ) - H_m^B(t') 
\right\}
\label{solhj}
\ee
Together with eqs. (\ref{hjact}, \ref{dst}), 
this expression shows that 
the energy term $-\Delta H_m dt$ originates
from the linear recoil of gravity, $\delta \pi_i dX^i$,
induced by the change in the matter trajectory.

Notice that all heavy degrees of freedom $X^i$
contribute to the determination of the background
time. In this 
we agree with what has been advocated 
in \cite{barbour} safe for the fact that the 
light variables only intervene in the definition of time
through their ``mean'' energy $H_m^B$. 
To our opinion 
however, the important lesson of eq. (\ref{hjt3}) is not 
that $t^B(X^i)$  is a ``redundant''
parameter 
but that it is precisely the one which does
the job: to parametrize the dynamical evolution
of the light variables that did not directly participate in its 
determination.
The reasons for which $t^B(X^i)$ is the right parameter
are the following. First, the change of the light trajectory
that we want to parametrize induces through the
constraint equation a change in the 
energy available to the heavy variables. 
Secondly $t^B(X^i)$ is defined by $-\partial_E S^B_X$,
i.e. the conjugate to an infinitesimal change of that
available energy.
Then to first order in the finite matter energy change,
these two reasons imply eq. (\ref{hjt3}).

Notice also that in spite of the fact that the new matter configuration
defines a new background orbit, the ``transversal'' part
of the change does not show up to first order in $\delta \pi_i$
since it is projected out by the contraction with the
gradient $C^{ij} \partial_i S^B \partial_j$ which is 
perpendicular to the surfaces $S^B_X = Const.$ 
This is a well known feature of first order
perturbation theory in classical mechanics, see e.g. \cite{mecacl}.

We now verify that the neglected term in eq. (\ref{hjt3}) 
governs back-reaction effects, i.e. 
the modifications of the matter dynamics
induced by its own energy through 
the ``susceptibility'' of the background.
To estimate 
this term perturbatively 
we use 
eq. (\ref{solhj}) and we obtain\footnote{
As pointed out to me by Serge Massar,
eq. (\ref{sus}) is exact only in the limit
of slowly varying $\Delta H_m (X^i)$, i.e.
when $\Delta H_m  \gg (X^i_2 - X^i_1) \partial_{X^i} \Delta H_m$.
This is the usual condition appearing in adiabatic
treatments\cite{arnold}.
Thus, for rapidly varying $\Delta H_m$ or
extremely long trajectories, 
one should exactly take into account 
the transversal dependence of $\Delta H_m (X^i)$
associated with the change in the end point value $X^i_2$.
This amounts to compute the Jacobi fields 
of $S^B_X$.
The reader interested by the exact calculation will
consult Section 5 of \cite{bk}. 
Let us mention here that it is necessary to 
perform the exact calculation only when the 
variables $X^i$ are not sufficiently ``heavy''.
Notice also that our estimate, eq. (\ref{sus}),
does not coincide with that of \cite{bk}. In that article,
the longitudinal is defined by the surfaces $S^B_X = Const.$
whereas here it is defined by $t^B= Const.$
A detailed comparison of these estimates goes beyond the 
scope of the present paper. 
For further discussion see next Section.}
\be
{ 1 \over 2}
C^{ij}(X^k) \partial_{X^i} \Delta S\; \partial_{X^j} \Delta S
=
{ 1 \over 2}
C^{ij}(X^k) \partial_{X^i} t^B \; \partial_{X^j} t^B (\Delta H_m)^2
\label{sus}
\ee
This term generalizes what we obtained in eq. (\ref{quadr}).
Indeed for backgrounds characterized by a single variable,
one has $C(X) \partial_{X} t^B  \partial_{X} t^B = 1/ C(X) \pi_X^2$
which is the integrand of the fourth expression in eq. (\ref{quadr}).
Moreover, in multi dimensional cases, 
when the change in energy $\Delta H_m$
is constant, 
eq. (\ref{sus}) still contributes
to $\Delta S$ given in 
eq. (\ref{solhj}) by the addition of $(\Delta H_m)^2 \partial_E t^B/2$.
By $\partial_E t^B$ we designate $\partial^2_E S^B_X$.
To evaluate it, one should vary eq. (\ref{prtl}) 
with respect to $d H_m^B=dE$.
One finds,
\ba
\partial_E
\left\{C^{ij}(X^k) \partial_{X^i} S_X^B \; \partial_{X^j} t^B(X^i) 
=
1 \right\} &&
\nonumber\\
C^{ij}(X^k) \partial_{X^i} t^B \; \partial_{X^j} t^B - 
C^{ij}(X^k) \partial_{X^i} S_X^B \; \partial_{X^j} \partial_E t^B
= 0 \ \ &&
\label{sus2}
\ea
Using once more eq. (\ref{prtl}) to rewrite the second term of
eq. (\ref{sus2}), one gets
\be
\partial_E t^B = \int^t_0 dt' C^{ij}(X^k) \partial_{X^i} t^B \; \partial_{X^j} t^B 
\label{fldis}
\ee
From this equation one verifies that
eq. (\ref{sus}) introduces a source term to 
eq. (\ref{hjt3}) which integrated over $t$ gives the announced
result in the cases of constant $\Delta H_m$.

When the last term of eq. (\ref{hjt2}) is replaced by the
r.h.s. of  eq. (\ref{sus}), one obtains an equation
for the matter propagation since all $X^i$ are
evaluated on their background values and parametrized by $t$. 
Thus one can view the last three terms of eq. (\ref{hjt2}) as defining
the ``effective'' matter hamiltonian. 
This hamiltonian
gives rise to unusual Euler-Lagrange equation since it is
quartic in $p_\phi= \partial_\phi \Delta S$. 
This results from 
the ``elimination'' of the degrees of freedom $X^i$.
A simple procedure to handle this hamiltonian consists in 
treating the correction term perturbatively, in a manner
similar to what is done in a Hartree approximation
and also in agreement to what we have learned in 
the former Section. Namely, the quadratic correction
term only shifts the saddle point energy $\bar \e$
by a term proportional to $\partial^2_\e S_G/ \partial^2_\e S_m$.

In this procedure, one first uses 
the unperturbed action eq.
(\ref{solhj}) to obtain the energy difference
$\bar \Delta H_m(t)$ evaluated along 
the unperturbed equations of motion.
Having obtained this function of $t$ (a c-number in the 
quantum mechanical language) one linearizes
eq. (\ref{hjt2}) with respect to $
H_m(\phi, \partial_\phi \Delta S, X^i(t) ) - H_m^B(t)$
so as to keep the quadraticity of the HJ equation in
$\partial_\phi \Delta S$. One gets
\be
\partial_t \Delta S =
\left[ 
H_m(\phi, \partial_\phi \Delta S, X^i(t)) - H_m^B(t)
\right]
\left[ 1 + {1 \over 2} \bar \Delta H_m(t) 
C^{ij} \partial_{i} t^B  \partial_{j} t^B
\right]
\label{modeq}
\ee
By construction, the correction term
is now expressed as a factor.
Moreover, since it depends only on $t$,
it can be absorbed in the l.h.s. by a redefinition of $t$.
Thus its sole effect 
is to modify the time-parametrization of the 
background orbit $X^{iB}(t)$ and therefore 
the lapse of time between the end points $X^i_1$
and $X^i_2$.
Notice that this procedure can be refined by 
determining self-consistently $\bar \Delta H_m(t)$.

\vskip .3 truecm

{\bf Application to quantum cosmology}

\noindent
The aim of this brief subsection is to show how
 similar are the quantized versions of the 
equations we just obtained.

As in the former Section, in order to extract the 
background contribution, we write the 
total kernel, solution of the WDW equation
(the quantized version of eq. (\ref{hjt})), as a product of the 
background gravitational propagator times the rest:
\be
{\cal{K}} ( X^i_2, \phi_2 ; X^i_1, \phi_1 )
= {\cal{K}}^B( X^i_2 ;  X^i_1) \times \Delta K( X^i_2, \phi_2 ; X^i_1, \phi_1 ) 
\label{factors}
\ee
The background gravitational 
 kernel is the solution of 
\be
\left[ - C^{ij}(X^k) \partial_{X^i}  \partial_{X^j}
+ V(X^k) + H_m^B(X^k)  
\right] {\cal{K}}^B( X^i_2 ;  X^i_1) =0
\label{nbkk}
\ee
Since the variables $X^i$ are heavy, it is 
legitimate to use the WKB approximation:
\be
 {\cal{K}}^B_{WKB}( X^i_2 ;  X^i_1) =
\sqrt{ Det. \partial_{X^\mu_2}  \partial_{X^\nu_1} S^B_X} \
\exp \left\{ i S^B_X(X^i_2) \right\}
\label{vv}
\ee
where the normalization factor is given by the enlarged
Van Vleck determinant which occurs when one works at fixed energy,
see \cite{shulm}. 
By $X^\mu_2$ and $X^\nu_1$, we designate the sets of variables $(X^i_2, dE)$
and $(X^j_1, dE)$.
(Notice that we should not worry about the 
normal ordering ambiguities of the operator 
$C^{ij}(X^k) \partial_{X^i}  \partial_{X^j}$ since we are working in the
WKB approximation.)

Using the fact that the background kernel is 
the WKB solution of eq. (\ref{nbkk}),
one immediately obtains the equation for $\Delta K$
\ba
&& \left[
C^{ij}(X^k) \partial_{X^i} S_X^B \; i\partial_{X^j} 
+ \hat H_m(\phi, i\partial_\phi , X^i ) - H_m^B(X^i ) 
\right] \Delta K( X^i, \phi; X^i_1, \phi_1)
\nonumber\\
&&\quad
+  {1 \over 2}\left[ -
C^{ij}(X^k) \partial_{X^i} \; \partial_{X^j} \right]
 {\sqrt{ Det. \partial_{X^\mu_2}  \partial_{X^\nu_1} S^B_X} \ 
\Delta K ( X^i, \phi; X^i_1, \phi_1) }=0
\label{hjt4}
\ea

The parallelism between the classical version eq. (\ref{hjt2}) 
and this equation is manifest.
In the quantum version, upon neglecting the
second line of eq. (\ref{hjt4}), i.e
to first order in small gradients $\partial_{X^i} \Delta K$ and
$\partial_{X^i}  Det.$,  $\Delta K$ is 
the usual matter kernel and it satisfies the Schr\"odinger equation
in the background geometry driven by $H_m^B(X^i)$.

Higher order terms
govern the corrections to the WKB and the BFA approximations. 
Their structure results from the {\it a priori} 
factorization of ${\cal{K}}^B_{WKB}$.
Thus one should be careful in interpreting 
these terms as providing {\it intrinsic} corrections 
to the WDW equation.
It should be indeed recalled that for unidimensional
mini-superspace models, the intrinsic corrections
have been determined
from the exact background-free 
equation governing matter evolution\cite{wdwpc}
which was obtained
without factorizing a given gravitational kernel 
nor postulating that the
WKB expressions constitute a good approximation. 
It was show that the intrinsic corrections
do not reproduce those resulting from 
a {\it a priori} factorization.
The challenging question is thus to extend this
analysis to the present multidimensional case.
We hope to be able to report on this
question.


\section { The background field approximation in cosmology}

In this Section we indicate how
the results of the former
one 
can be generalize
to arbitrary three-geometries whose metric
is given by $g^{ij}(x)$.
The total hamiltonian which governs the entangled dynamics of
 $g^{ij}(x)$ and some local matter field $\phi(x)$ has the following
structure
\be
H_T =
\int d^3x N^\mu(\xi, x) {\cal{H}}_\mu^T(\xi, x) = 
\int d^3x N^\mu(\xi, x) \left\{ {\cal{H}}_\mu^G(\pi_{ij}, g^{ij} ) 
+ {\cal{H}}_\mu^m(p_\phi, \phi, g^{ij})
\right\} 
\label{gham}
\ee
The four-vector field $N^\mu(\xi, x)$ describes how successive
3-surfaces are pasted to form a 4-dimensional space-time,
see \cite{mtw}. ${\cal{H}}_\mu^G(\xi, x)$
is the gravitational energy-momentum density. 
It generalizes eq. (\ref{hg}), it 
is quadratic in the gradients $\partial_{x^k} g^{ij}(x)$
and in the conjugate momenta $\pi_{ij}(x)$
but is a non-linear function of $g^{ij}(x)$.
The matter energy-momentum vector density
${\cal{H}}_\mu^m(\xi, x) $ depends on $\phi(x)$,
$\partial_{x^k} \phi(x)$, $p_\phi(x)$ and $g^{ij}(x)$.

The main difference with Section 3 is that there are 
now four local equations of constraint in the place of 
eq. (\ref{hjt}) which result from the four Einstein's equations
$\partial_{N^\mu} S_T =  {\cal{H}}_\mu^T(\xi, x) =0$.
In the HJ guise, they read
\ba
{ 1 \over 2}
C^{iji'\!j'\!}(g^{kl}(x)) \partial_{g^{ij}} S_T \; \partial_{g^{i'\!j'\!}} S_T + V(g^{kl}(x))
+ {\cal{H}}_0^m(\phi(x), \partial_{\phi} S_T, g^{ij}(x)) =0&& \quad
\label{hjtloc1}
\\
 - 2 ( \partial_{g^{ij}} S_T)^{\vert j} + 
{\cal{H}}_i^m (\phi(x), \partial_{\phi} S_T, g^{ij}(x)) =0 &&
\label{hjtloc}
\ea
For further details concerning the kinetical matrix $C$ or the
potential $V$ see \cite{mtw}.
The important point is that these constraints do not decouple since
the potential term  $V(g^{kl}(x))$ contains gradients
connecting the fields at different points. Upon neglecting this
term one obtains an action which is just a sum of decoupled
actions evaluated at each point. 
As shown in \cite{ss},
one can then perform a perturbative treatment in 
those gradients. What follows can be conceived as
a generalization of these works in that the small parameter 
which governs the expansion is 
the change of the matter hamiltonian itself.
Therefore, our (less explicit) treatment also applies to arbitrary
large gradients. 

In spite of these gradients, 
the total action from the initial configuration
{$g^{ij}_1(x),\phi_1(x)$} to the final one {$g^{ij}_2(x),\phi_2(x)$}
which is a solution of the constraints can
be decomposed, as in eq. (\ref{hjact}), as
\be
S_T = \int^{g_2^{ij}(x)}_{g^{ij}_1(x)} dg^{ij}(x) \; \pi_{ij}(g^{ij}; \{ g_2^{kl},
\phi_2\} )+ \int^{\phi_2(x)}_{\phi_1(x)} d\phi(x)  p_\phi( \phi; \{ g_2^{kl},
\phi_2\})
\label{lochjt3}
\ee
where the local on-shell momenta are parametrized by their
conjugate field.
In this paper, we shall not discuss the difficulties that might
prevent to actually compute $S_T$ when the initial and
final configurations are given, i.e. to solve the ``thick''
sandwich problem. Rather we shall only use the structure of $S_T$
and the fact that it is a solution of eqs. (\ref{hjtloc1}, \ref{hjtloc})
since it is all what we need in our implementation 
of a BFA to gravity.

To conform ourselves to the ``standard procedure''\cite{kiefer, bk, isham}
(later on we shall comment on its appropriate character),
we treat all $g^{ij}(x)$ in the same footing as ``heavy''
 degrees of freedom. Thus, as in the former Section,
we choose an initial
matter configuration ($p_{\phi 1}(x), \phi_1(x)$) which determines
both the background matter density ${\cal{H}}_\mu^B$
and the background trajectories.
Then, as in eq. (\ref{dst}), we decompose $S_T$ as
$S_G^B(g^{ij}_2) + \Delta S (\phi_2, g^{ij}_2)$
and we perform an expansion in the 
changes $\delta \pi_{ij}(x)$ which result from the change in the
matter field configuration induced by the replacement
of the background value $\phi^B(x)$ by the new field
configuration specified by the final value $\phi_2(x)$.
By inserting this decomposition in eqs. (\ref{hjtloc1}, \ref{hjtloc}), 
we obtain respectively
\ba
&&
C^{iji'\!j'\!}(g^{kl}(x)) \partial_{g^{ij}} S^B_X \; \partial_{g^{i'\!j'\!}}\Delta S 
+ {\cal{H}}_0(\phi(x), \partial_{\phi} \Delta S, g^{ij}(x)) -  {\cal{H}}_0^B(x, g^{ij}(x)) 
\nonumber\\
&&
\quad \quad
+ {1 \over 2}C^{iji'\!j'\!}(g^{kl}(x)) \partial_{g^{ij}} \Delta S 
\; \partial_{g^{i'\!j'\!}} \Delta S  =0
\label{lochjt4}
\\
&& 
- 2 ( \partial_{g^{ij}} \Delta S)^{\vert j} + 
\left(
{\cal{H}}_i^m (\phi(x), \partial_{\phi} S_T, g^{ij}) - {\cal{H}}_i^{mB}(x, g^{ij}(x))
\right) = 0
\label{ndeltaloc}
\ea

In order to relate these local equations to a
single HJ equation governing matter propagation in 
a given background metric,
we must introduce by hand the lapse-shift vector field $N^\mu(x, t)$.
These four functions then
define univocally a single time parameter.
It is given by $t = - \partial_E S^B_X$
where
$dE$ is an infinitesimal constant 
change in the total matter energy  $\int d^3 x N^\mu(x){\cal{H}}^{m B}_\mu(x)$. 
More precisely, this ``redundant'' parameter satisfies
\be
\int d^3 x \left\{ N^0(x) 
C^{i'\!j'\!ij}(g^{kl}(x)) \partial_{g^{i'\!j'\!}} S^B_X \; \partial_{g^{ij}} 
+ 2 N^i(x)^{\vert j} \partial_{g^{ij}}
\right\} t^B(g^{ij}(x), N^\mu(x))= 1
\label{lochjt5}
\ee
This equation 
is obtained by varying eq. (\ref{gham}) with respect to $dE$ 
and by imposing $H_T =0$.
It allows to recover 
the usual relation 
between the momentum and the velocity:
 $dg^{ij}(x)/dt = N^0 C^{iji'\!j'\!} \pi_{i'\!j'\!}
+ 2 N^{i\vert j}$.

Using this time parameter
and eq. (\ref{ndeltaloc}), and upon neglecting the
second line of eq. (\ref{lochjt4}) which is quadratic in the 
small gradients $\partial_{g^{ij}} \Delta S$,
the first line of that equation
becomes the HJ equation
\be
\partial_t \Delta S = \int d^3 x N^\mu(t, x) \left\{ {\cal{H}}_\mu^m(
\phi(t, x), \partial_{ \phi(t, x)} \Delta S , g^{ijB}(t, x)) 
- {\cal{H}}_\mu^{mB}(t, x) 
\right\}
\label{newlochj}
\ee
evaluated in the curved geometry solution of the
Einstein's equation driven by 
the ``mean'' matter energy density ${\cal{H}}_\mu^{mB}(x)$.

In brief, the novelty introduced by considering arbitrary
3-geometries relies on the necessity of introducing
from the outset the four functions $N^\mu(x)$ in order
to rigidify the propagation in super-space from the initial
 3-geometry to the next ones.
We emphasize that the dynamical justification arises from
the presence of gradients $\partial_x g^{ij}, \partial_x \phi$
in the potential terms of $H_T$.


In what follows, we shall not present the quantum version 
of this derivation which would lead to the Schr\"odinger 
equation in a curved geometry. 
It is straightforwardly obtained by generalizing 
eqs. (\ref{factors}-\ref{hjt4}). 
(The interested reader will find this quantum treatment in \cite{bk}
which is however restricted to empty solutions of Einstein equations,
i.e. to ${\cal{H}}^B_{m \mu} = 0$.)
Once more, the reader will verify that the structure of 
the quantum version is fully determined by
 the implementation of a BFA in classical settings.

We shall neither present the computation of the quadratic
corrections 
which arise through the second line of eq. (\ref{lochjt4}).
They can also be found in \cite{bk}.
What we shall do instead is to discuss
on physical grounds the appropriate 
character of applying, as we just did, a BFA
to all gravitational degrees of freedom.

We shall base our discussion on the
fact that the quadratic corrections to $\Delta S$ are 
given by a term of the form $\int \!\!\int T_{\mu \nu} 
D^{\mu \nu \mu'\!\nu'} T_{\mu'\! \nu'}$
where $D$ is a gravitational Green function.
These interactions are in strict analogy with the current-current 
interactions in electro-magnetism obtained by integrating over 
the photon field. 
However, as discussed in \cite{bk},
  the Green function which is obtained 
by treating all gravitational degrees of freedom
on the same footing is rather 
unconventional in that it vanishes both on the initial and 
final 3-surfaces, in contradistinction with the usual 
retarded and Feynman functions in classical and 
quantum settings. This unconventional feature
results from the fact that the $g^{ij}$ have been completely 
specified on both the initial and final 3-surfaces. 

This unpleasant feature reveals that the ``heavy'' character 
that has been {\it a priori} 
attributed to $g^{ij}$ must be called into question.
Indeed, so far, $g^{ij}$ has been treated like the $W$ gauge boson
in the derivation of the four fermions weak interactions model.
In that case, it is the rest mass of the $W$ which legitimizes 
its elimination for low energy processes since no on-shell
$W$ is ever produced.
For gravity, the Planck mass does not act
like the $W$ mass and no mass threshold prevents 
the production of on-shell gravitons. For this 
reason, it is inappropriate 
to functionally integrate out the gravitons. 
Indeed, in quantum gravity, these gravitons are
correlated in phase and amplitude to the quantized matter
sources. This quantum entanglement prevents
the possibility of describing the final gravitational
configurations by $c$-number functions giving rise to 
a classical background action.
More intuitively, on which physical basis 
could one justify an unsymmetrical 
treatment in which rare (i.e. not coherent) on-mass gravitons 
are considered ``heavy'' and thus described by WKB 
waves on which one applies a BFA 
and photons with the same frequency are ``light'' and thus fully quantized?

In order to treat gravitons 
and photons on an equal footing, one must therefore abandon the
idealized procedure in which all gravitational degrees of freedom
are treated on the same footing. By treating the 
light part of the gravitational degrees of freedom like
matter degrees of freedom, one ``recovers''
what has been adopted in cosmology to study the growth of 
perturbations\cite{HH}, see also \cite{Hal}
for a discussion on the dynamical role of light and 
fluctuating gravitational degrees of freedom in explaining why 
the heavy coordinates are driven by mean values.

In conclusion, it is physically unjustified to search
for a treatment of the solutions of the WDW in which all
gravitational degrees of freedom are treated on the same footing.
However what is lost from a geometrical point of view is compensate
by a dynamical justification: in each situation, it will be
convenient and appropriate to treat some gravitational 
degrees of freedom as heavy but it will also be
mandatory to keep 
the other ones fully quantized. The same division 
also applies to matter degrees of freedom.
The implementation of this program requires
further work which will hopefully
be presented somewhere else.

\vskip .3 truecm\vskip .3 truecm
{\bf Ackowledgements}

I am indebted  to Serge Massar for many 
fruitful discussions and remarks.
I also wish to thank Bruno Boisseau and
Amaury Mouchet for useful suggestions 
concerning aspects of the HJ formulation of
classical mechanics.

\vskip .3 truecm

\section { Appendix A. $ $
The background field approximation based on trajectories.}
 
Given the hamiltonian eq. (\ref{TA0}),
the total ``off-shell'' action we are considering is
\be
S_T = \int_0^1 d\xi \left\{
\pi_i \dot X^i + p_\phi \dot \phi - N(\xi) 
\left[ H_X(\pi_i, X^i) + H_m(\phi, p_\phi, a )
\right]\right\}
\label{TA}
\ee
In order to have univocally specified expressions,
we must choose a ``representative'' for $N(\xi)$.
Indeed, at fixed end point values $X^i_1, \phi_1$
to $X^2_2, \phi_2$, the action $S_T$ is invariant 
under reparametrizations of $\xi$ which
leave the end point values $\xi = 0, 1$ fixed.
The reparametrizations which satisfy this restriction
leave the integral
$\int_0^1 d\xi N(\xi)= N$ invariant. Thus one can only  
(but always) choose as a representative the constant 
lapse $N(\xi) = N$. This choice is the one usually made
when describing the path integral of a relativistic
particle. In that case, $ds= N d\xi$ defines
the Schwinger fifth-time. Adopting the same
choice, our ``representative'' off-shell action is
\be
S_T = \int_0^N dt \left\{
\pi_i \dot X^i + p_\phi \dot \phi -  
\left[ H_X(\pi_i, X^i) + H_m(\phi, p_\phi, X^i)
\right]\right\}
\label{reTA}
\ee
We have used the symbol $t$ defined by $dt = N d\xi$,
since in this ``gauge'' of constant lapse,
it indeed corresponds to the proper time.
In this gauge,
the total hamiltonian is a constant along the classical 
trajectory which relates $X^i_1, \phi_1$
to $X^i_2, \phi_2$ in a proper time $N$ since 
it is not explicitly time dependent.

Since physical propagation must occur with
vanishing total energy, one must impose
$\partial_N S_T = H_X + H_m = 0$.
This extremization condition determines the 
``saddle value'' $\bar N=N(X^i_2, \phi_2; X^i_1, \phi_1)$
which,
in our gauge, equals $t(X^i_2, \phi_2)$,
the propre time lapse 
from $X^i_1$ to $X^i_2$ given $\phi_1$ and $\phi_2$.
 $\partial_N S_T =0$  plays thus the role 
of $\partial_\e S_T = 0$
in Section 2: Both fix the proper time
lapse on the classical trajectory.

Thus the fully extremized action is
\be
S_T = S_X + S_m = \int_0^{\bar N} dt \left\{
\pi_i \dot X^i + p_\phi \dot \phi
\right\}
\label{stex}
\ee
In this action, $X^i(t)$ and $\phi(t)$ are evaluated along the
equation on motion, they thus depend parametrically on $\bar N$.

We are now in position to implement a BFA.
As in Section 2, we shall linearize only the gravitational 
part of the action in order to obtain the simplified description of the matter evolution for 
values of $\phi_2$ centered around the background value $\phi_2^B$.
The difference with that Section is that we shall consider
the difference of extremized actions since no longer can work with
$\e$ off shell. 

Thus, we want to expand
$S_T(X^i_2, \phi_2, \bar N(X^i_2, \phi_2))$
around the background action $S_T(X^i_2, \phi^B_2, 
\bar N^B= \bar N(X^i_2, \phi^B_2))$
to a first order change of the background coordinates $
\delta X^i(t)$ 
and the lapse $\delta N$.
We proceed in three steps. 

In the first step, we replace $\bar N(X^i_2, \phi_2)$ by its background
value $\bar N^B= \bar N(X^i_2, \phi^B_2)$ in the first action.
To first order in $\delta N$, the action is left 
unchanged since we are working with vanishing total energy
and we are not modifying the end point values of $\phi$ and $a$.
To consider only the first order change in $\delta N$ is a valid
restriction
because, when $X^i(t)$ are heavy, i.e. when second order change 
in $\delta X^i(t)$ are negligible with respect to first order change, 
the same in true for the changes due to $\delta N$. In other words
when $X^i(t)$ are heavy, $N$ is also heavy.
(This is exactly like in statistical mechanics. The change
of the temperature induced by a fluctuation of the 
energy repartition scale like the inverse volume 
of the heavy system.)

In a second step, we consider the first order change in $\delta X^i(t)$
and $\delta \pi_i(t)$. We have
\be
S_X(X^i_2, \phi_2, \bar N^B) = S_X(X^i_2, \phi^B_2, \bar N^B)
+ \int_0^{\bar N^B} dt \left\{\delta \pi_i \dot X^i - \delta X^i \dot \pi_i
\right\}
+ \delta X^i \pi_i \vert^{X^i_2}_{X^i_1}
\label{2ste}
\ee
Since we are working with fixed end points $X^i_1, X^i_2$,
the boundary term vanishes.
Moreover since we are on-shell,
\be
\delta \pi_i \dot X^i - \delta X^i \dot \pi_i = \delta_{X^i, \pi_i}
(H_X + H_m)
\label{dapi}
\ee
where $\delta_{X^i, \pi_i}$ means the variation of the
quantity induced by
the sole variations of $X^i$ and $\pi_i$, i.e. with $\phi(t)$
and $p_\phi(t)$ evaluated along their background classical 
values.

In the third step, we re-use the vanishing of the total energy
for both 
classical trajectories. 
Thus
\ba
\delta_{X^i, \pi_i}
(H_X + H_m) &=& - \delta_{\phi, p_\phi}(H_X+ H_m) 
= - \delta_{\phi, p_\phi} H_m
\nonumber\\
&=&
- H_m( p_\phi (t), \phi(t), X^{iB}(t))+
 H_m( p^B_\phi (t), \phi^B(t), X^{iB}(t))
\label{dHdH}
\ea
The second equality follows from the fact that $H_X$ is independent of $
\phi$ and $p_\phi$. The last one
means that the variations with respect to $\phi$ and $p_\phi$ 
should be taken to all orders and not to first order only
in the case of $\delta X^i$ and $\delta \pi_i$.
We emphasize that both terms in this last expression
are evaluated in the background geometry $X^{iB}(t)$.

By collecting the results, to first order in $\delta N,
\delta X^i, \delta \pi_i$, we have
\ba
S_T(X^i_2, \phi_2, \bar N(X^i_2, \phi_2)) &=& S_X(X^i_2, \phi^B_2, \bar N^B)
+ \int_0^{\bar N^B} dt H_m^B(t)
\nonumber\\
&&\quad
+ \int_0^{\bar N^B} dt 
\left\{ p_\phi (t) \dot \phi(t) - 
H_m( p_\phi (t), \phi(t), X^{iB}(t))
\right\} \quad
\label{fov}
\ea
The $\phi_2$ dependent part of this action is the usual
HJ matter action evaluated in the time dependent
background $X^{iB}(t)$.
The simplest way to verify it is to consider
first order variation of the end point values $X^i_2$.
At this point indeed, we cannot vary the time lapse $\bar N^B$
since it is fixed by $a_2$. The first order variation of
$S_T$ with respect to an arbitrary variation of the
end points $X^i_2$ gives
\be
\delta S_T = \delta X^i_2 \;
\partial_{X^i_2} S_T = \delta X^i_2
\left\{ g_i(X^j_2) + 
 \partial_{X^i_2} \bar N^B(X^i_2, \phi_2) \; 
H_m( \partial_\phi S_T, \phi, X^i_2)
\right\}
\label{hja}
\ee
where the dependence in $X^i_2$
which is irrelevant
for the matter dynamics has been put in the functions $g_i$. 
Using $\bar N^B = t(X^i_2)$, we can now determine 
the change of $S_T$
when  
the $\delta X^i_2$
are evaluated along the classical trajectory, i.e.
when they are of the form $\delta X^i_2 = C^{ij} \partial_j S_X^B \delta t$.
By definition of $t$ one obtains
\be
\delta S_T = \delta t \; \partial_t S_T = \delta t 
\left\{H_m( \partial_\phi S_T , \phi, a^B(t))
 + g_i(t) \partial_t X^{iB}(t)
\right\}
\label{hjeq}
\ee
It is thus through changes of $X^i$ 
evaluated along the background orbit that
one recuperates the time dependent
HJ equation for the matter degrees of freedom. 

\section { Appendix B. $ $The gravitational response to
off-shell matter configurations}
 
In the former Appendix we saw that the linear change of the 
background action $S_X$ induced by the change of the
matter configurations is equal to $\int dt \Delta H_m$ evaluated
along the background trajectory $X^{iB}(t)$, see eq. (\ref{fov}).
However it should be noticed that
this result has been obtained by comparing neighboring
{\it on-shell} matter configurations. 
It would therefore be very useful to extend this result
to off-shell matter configurations in order to be able to
use this relationship in a path integral formulation of
quantum mechanical matter propagation 
which does involve off-shell configurations.

Thus the problem one should analyze in classical settings
is to determine the gravitational response to a given 
arbitrary matter configuration $\phi(\xi), p_\phi(\xi)$.
Strictly speaking, this cannot be achieved 
since one cannot find a stationary gravitational
action whose source would be this particular matter configuration.
Indeed, there is no reason that this configuration satisfies
the conservation equation eq. (\ref{bianch}). Therefore, the Bianchi
identity tells us that no (on-shell) background could be driven by this source.

However this does not prevent us to characterize the {\it
departure} from stationarity when the configuration $\phi(\xi), p_\phi(\xi)$
is close to a classical orbit $\phi^B(\xi), p^B_\phi(\xi)$.
As we shall see, this departure is just what we need to
recover the usual off-shell matter action  in a given background
that one uses in a path integral formulation.
To prove this result, we shall simply adapt the procedure
we used in the former Appendix so as to take into account the 
off-shell character of the configuration $\phi(\xi), p_\phi(\xi)$.

The total unextremized action is given by eq. (\ref{reTA}) 
since one can still work with constant lapses $N$ as
representatives. (The off-shellness of $\phi(\xi)$ does not spoil
reparametrization invariance). Therefore, at fixed
$N$, one can still extremize $S_T$ with respect
to arbitrary variations of $X^i(t)$ and $\pi_i(t)$
with fixed end points $(X^i_1, X^i_2)$.
Doing so one obtains an ``almost'' extremized action
since it is stationary for all variations but those of the 
constant lapse $N$. 

At this point the ``proximity'' of $\phi(t), p_\phi(t)$
with some classical trajectory $\phi^B(t), p^B_\phi(t)$
must be called upon.
Indeed, to first order in the change in the lapse around
its background value $\bar N^B$, one can
replace $N$ in the ``almost'' extremized action 
by the background value $\bar N^B$
and obtain
\be
S_T = \int_0^{\bar N^B} dt \left\{
\pi_i \dot X^i + p_\phi \dot \phi -
\left[ H_G(X^i(t), \pi_i(t)) + H_m (\phi, p_\phi, X^i(t)) \right]
\right\}
\label{stex2}
\ee
where $X^i(t)$ is evaluated along the stationary orbit 
from $(X^i_1, X^i_2)$ in a time $\bar N^B$ which is driven 
by the off-shell ``pressures'' $\partial_{X^i} H_m(\phi, p_\phi, X^i)$.
The reason why this replacement can be performed is the following.
Upon searching $\bar N$, the solution of $\partial_N S_T=0$ where $S_T(N)$ 
is the ``almost'' extremized action defined above, one finds
that $\bar N$ is complex function since no classical (i.e. real)
solution can be found. However, when $\phi(t), p_\phi(t)$ is close
to $\phi^B(t), p^B_\phi(t)$, i.e. when  $H_m(\phi, p_\phi) -
 H_m(\phi^B, p^B_\phi) \ll H_m(\phi^B, p^B_\phi)$ for all $t$,
 the real part of $\bar N$ is close to 
$\bar N^B$ and that its imaginary part is small.
Remember that the location of the
saddle is mainly determined by the heavy coordinates $X^i$.
Thus, one can correctly replace the complex $\bar N$ 
by the background value $\bar N^B$.
Then, the off-shellness of the matter configuration
is made manifest by the fact the $H_G + H_m $ in eq. (\ref{stex2})
does not vanishes. Not surprisingly, it is through this term that the usual
matter action will be recovered.

Having performed this first step, one can indeed proceed
to the second and third step of Appendix A.
The simplest way to proceed consists in ``adding''
$H_G(X^{iB}, \pi_i^B) + H_m (\phi^B, p_\phi^B, X^B) = 0$
to the integrand of eq. (\ref{stex2}) and implementing directly
eqs.  (\ref{2ste}) and (\ref{dapi}) in the resulting action.
(Both equations
rely only on the
stationary character of the orbits $X^i(t)$ and $X^{iB}(t)$.)
After some elementary algebra, one recovers eq. (\ref{fov}).

What we have learned from this generalization is that 
the determination of the {\it linear} gravitational response 
does not require on mass-shell matter sources.
Thus the matter action which determines 
the phase of off-shell configurations in a 
path integral formulation of quantum matter propagation in a 
given background can be considered as the action 
emerging, through a first order variation,
from a more quantum framework in which 
gravity responded to these off-shell configurations.
Even though the imaginary part of $\bar N$
is negligible for heavy coordinates $X^i$, strictly speaking
this imaginary part is present. So what does it mean?
Simply that in the path integral of
quantum gravity, these gravito-matter configurations
 will be exponentially reduced by this imaginary 
contribution with respect to the BFA formalism in which 
one has discarded this contribution.

\end{document}